\documentclass[12pt,a4paper,final]{iopart}

\usepackage[english]{babel}
\usepackage{iopams}
\usepackage{cite}

\expandafter\let\csname equation*\endcsname\relax

\expandafter\let\csname endequation*\endcsname\relax

\usepackage[utf8]{inputenc}

\usepackage{amsmath}
\usepackage{iopams}
\usepackage{graphicx}
\usepackage{color}
\usepackage[breaklinks=true,colorlinks=true,linkcolor=blue,urlcolor=blue,citecolor=blue]{hyperref}

\newcommand{\NN}{\mathcal{N}}
\newcommand{\TT}{\mathcal{T}}
\newcommand{\cc}{\mathrm{cc}}
\renewcommand{\ss}{\mathrm{ss}}
\renewcommand{\sc}{\mathrm{sc}}
\newcommand{\cs}{\mathrm{cs}}
\newcommand{\typ}{\mathrm{typ}}
\newcommand{\D}{\mathrm{d}}
\newcommand{\E}{\mathrm{e}}
\newcommand{\I}{\mathrm{i}}
\newcommand{\abs}[1]{\left\lvert{#1}\right\rvert}

\newcommand{\var}{\operatorname{var}}

\begin{document}


\title[Noise-to-signal ratio]{Noise-to-signal ratio of single-trajectory spectral densities in centered Gaussian processes}

\author{Alessio Squarcini$^{1,2,3}$, Enzo Marinari$^{4,5}$, Gleb Oshanin$^6$, Luca Peliti$^7$ \& Lamberto Rondoni$^{8,9}$}
\address{$^1$ Institut f\"ur Theoretische Physik, Universit\"at Innsbruck, 
Technikerstrasse 21A,
A-6020 Innsbruck, Austria\\
$^2$ Max-Planck-Institut f\"ur Intelligente Systeme, Heisenbergstrasse 3, D-70569, Stuttgart,
Germany\\
$^3$ IV. Institut f\"ur Theoretische Physik, Universit\"at Stuttgart, Pfaffenwaldring 57,
D-70569 Stuttgart, Germany\\
$^4$ Dipartimento di Fisica, Sapienza Universit{\`a} di Roma, P.le A.
Moro 2, I-00185 Roma, Italy\\
$^5$ INFN, Sezione di Roma 1 and Nanotech-CNR, UOS di Roma, P.le A.
Moro 2, I-00185 Roma, Italy\\
$^6$ Sorbonne Universit\'e, CNRS, Laboratoire de Physique Th\'eorique de la Mati\`{e}re Condens\'ee (UMR CNRS 7600), 4 place Jussieu,
75252 Paris Cedex 05, France\\
$^7$ Santa Marinella Research Institute, Santa Marinella, Italy\\
$^8$ Dipartimento di Scienze Matematiche, Politecnico di Torino, Corso Duca degli Abruzzi 24, 10129 Torino, Italy\\
$^9$ INFN, Sezione di Torino, Via P. Giuria $1$, 10125 Torino, Italy}

\begin{abstract} 
We discuss the statistical properties of a 
single-trajectory power spectral density $S(\omega,\TT)$ of an arbitrary real-valued centered Gaussian process $X(t)$, where $\omega$ is the angular frequency and $\TT$ the observation time.
We derive a double-sided inequality for its noise-to-signal ratio and obtain the full probability density function of $S(\omega,\TT)$. Our findings imply that the fluctuations of $S(\omega,\TT)$ exceed its average value $\mu(\omega,\TT)$. This implies that using $\mu(\omega,\TT)$ to describe the behavior of these processes can be problematic. We finally evaluate the typical behavior of $S(\omega,\TT)$ and find that it deviates markedly from the average $\mu(\omega,\TT)$ in most cases.
\end{abstract}

Keywords: Single-trajectory power spectral density, Gaussian stochastic processes, Noise-to-signal ratio

\maketitle

\section{Introduction}

The power spectral density (PSD) of a deterministic or stochastic process $X(t)$ encodes important information about its properties and is widely used in experimental, numerical and theoretical analyses (see, e.g., refs.~\cite{norton,flandrin,eli,a} and references therein). The single-trajectory PSD $S(\omega,T)$ is defined by
\begin{align}
\label{def1}
S(\omega,\TT) = \frac{1}{\TT} \abs{\int_0^\TT \D t \; \E^{\I \omega t} \, X(t)}^2 \,,
\end{align}
where $\omega$ is the (angular) frequency and $\TT$  the 
observation time. Usually, one averages $S(\omega,\TT)$ over an \textit{ensemble} of trajectories and eventually takes the $T \to \infty$ limit to get
\begin{align}
\label{def2}
\mu(\omega) =  \lim_{\TT\to \infty} \mu(\omega,\TT)\,, \qquad \mu(\omega,\TT) = \overline{S(\omega,\TT)} \,,
\end{align}
where the overbar here and henceforth denotes the ensemble averaging. We emphasise that $\mu(\omega)$ in eq.~\eqref{def2}---an ensemble-averaged property taken in the limit $\TT\to \infty$---is conventionally referred to as the PSD and its calculation is the usual target of the standard analyses of spectral properties of random processes. 

Let us remark that the $\TT \to \infty$ can be formally taken in mathematical expressions but not in experimental or numerical analyses, and that therefore caution is required when making comparisons with theoretical predictions. Moreover, for many 
non-stationary stochastic processes the infinite-$\TT$ limit of the expressions in eqs.~\eqref{def1} and \eqref{def2} does not exist, what requires either using some alternative definitions of power spectral densities (see, e.g., \cite{flandrin,alessio}) or to confine oneself to the finite-$\TT$ behaviour. Lastly, it is not always possible to have a large enough statistical sample in order to reliably perform the averaging. 

Motivated by the latter circumstance, recent works \cite{krapf1,krapf2,sposini1,sposini2,sara} have concentrated on stochastic properties of random variable $S(\omega,\TT)$ defined in eq.~\eqref{def1}.  
Indeed, it is important to know how $S(\omega,\TT)$ fluctuates from sample to sample, in order to estimate how large a statistical sample should be to allow a reliable evaluation of $\mu(\omega,T)$ from experimental or numerical analyses, especially in view of taking its large~$\TT$ limit.

  
It turns out that, for several centered Gaussian processes---standard Brownian motion \cite{krapf1}, fractional Brownian motion with Hurst index $H$ \cite{krapf2}, scaled Brownian motion \cite{sposini1}, diffusing diffusivity processes \cite{sposini2} and the Brownian gyrator model \cite{sara}---the probability density function $P(s)=P\left(S(\omega,\TT) = s\right)$ of the random variable $S(\omega,\TT)$ is explicitly given, for arbitrary values of~$\omega$ and~$\TT$, by the universal form
\begin{equation}\label{prob}
P(s) =  \frac{1}{\NN}\, \exp\left(- \frac{s}{(2 - \gamma^2) \mu(\omega,\TT)}\right) I_0\left(\frac{\sqrt{\gamma^2 - 1}}{2 - \gamma^2} \frac{s}{\mu(\omega,\TT)}\right) \,,
\end{equation}
where
\begin{equation}
\NN = \sqrt{2-\gamma^2} \mu(\omega,\TT) \,,
\end{equation}
$I_0(z)$ is the modified Bessel function and $\gamma$ is the ``noise-to-signal'' ratio, defined by
\begin{align}
\label{gamma}
\gamma = \gamma(\omega,\TT) = \frac{\sqrt{\overline{S^2(\omega,\TT)} - \mu^2(\omega,\TT)}}{\mu(\omega,\TT)} \,.
\end{align}
This latter parameter, which is also called the coefficient of variation of the distribution~\eqref{prob},
is a measure of the relative weight of fluctuations of the finite-$\TT$ PSD around its mean value. Note that the form of eq.~\eqref{prob} requires $\gamma$ to satisfy the double-sided inequality
\begin{align}
\label{ineq}
1 \leq \gamma \leq \sqrt{2} \,,
\end{align}
which was directly verified in ~\cite{krapf1,krapf2,sposini1,sposini2,sara} for each particular case under study.

Because the inequality in eq.~\eqref{ineq} and the distribution \eqref{prob}
appear to be valid  for rather diverse Gaussian stochastic processes, one can conjecture that they hold in general for an arbitrary centered Gaussian process. In this paper we focus on this question and present a formal proof that this is indeed the case.

In Section~\ref{1} we 
prove the crucial inequality~\eqref{ineq} for an arbitrary 
Gaussian centered process with arbitrary $\omega$ and $\TT$. Our proof is solely based on Wick's theorem. 
In Section~\ref{2} we take advantage of the Karhunen-Loeve decomposition of an arbitrary Gaussian process to calculate the moment-generating function of $S(\omega,\TT)$, from which the general result in eq.~\eqref{prob} follows by the inversion of the Laplace transform. Section~\ref{4} is devoted to the analysis of the functional forms of $P(s)$ in two limiting cases, and also to a discussion of the \textit{typical} behavior of $S(\omega,\TT)$.
We conclude with a brief summary of our results in Sec.~\ref{3}.

\section{The noise-to-signal inequality}
\label{1}

The second moment $\overline{S^2(\omega,\TT)}$ 
of a single-trajectory PSD of a real-valued process $X(t)$ can be formally written down as
\begin{equation}
\label{zz}
\begin{split}
\overline{S^2(\omega,\TT)} &= \frac{1}{T^2}  \int^\TT_0   \int^\TT_0  \int^\TT_0   \int^\TT_0\D t_1\, \D t_2\, \D t_3\, \D t_4 \; \overline{X(t_1) X(t_2) X(t_3) X(t_4)} \\
& \qquad {}\times \cos(\omega(t_1-t_2)) \cos(\omega(t_3 - t_4)) \,.
\end{split}
\end{equation}
According to Wick's theorem, the four-time  correlation function of the form $\overline{X(t_1) X(t_2) X(t_3) X(t_4)} $ for an arbitrary centered Gaussian process naturally decomposes as follows:
\begin{equation}
\begin{split}
\overline{X(t_1) X(t_2) X(t_3) X(t_4)}  &= \overline{X(t_1) X(t_2)}  \, \overline{X(t_3) X(t_4)} +  \overline{X(t_1) X(t_3)}  \, \overline{X(t_2) X(t_4)} \\ 
&\qquad {}+ \overline{X(t_1) X(t_4)} \,  \overline{X(t_2) X(t_3)} \,.
\end{split}
\end{equation}
This implies that the expression \eqref{zz} can be formally rewritten as
\begin{align}
\label{k}
\overline{S^2(\omega,\TT)} &= \mu^2(\omega,\TT) + 2 \overline{S_\cc(\omega,\TT)}^2 + 2 \overline{S_\ss(\omega,\TT)}^2 + 4 \overline{S_{\cs}(f,\TT)}^2 \,,
\end{align}
where we have used the shorthand notations
\begin{equation}
\begin{split}
\overline{S_\cc(\omega,\TT)} &= \frac{1}{\TT} \int^\TT_0 \int^\TT_0 \D t_1\, \D t_2 \; \overline{X(t_1) X(t_2)} \cos(\omega t_1) \cos(\omega t_2) \,, \\
\overline{S_\ss(\omega,\TT)} & = \frac{1}{\TT}  \int^\TT_0 \int^\TT_0 \D t_1\, \D t_2 \; \overline{X(t_1) X(t_2)} \sin(\omega t_1) \sin(\omega t_2) \,,\\
\overline{S_{\cs}(\omega,\TT)} & = \frac{1}{\TT}  \int^\TT_0 \int^\TT_0 \D t_1\, \D t_2 \; \overline{X(t_1) X(t_2)} \cos(\omega t_1) \sin(\omega t_2) \\ 
&= \frac{1}{\TT}  \int^\TT_0 \int^\TT_0 \D t_1\, \D t_2 \; \overline{X(t_1) X(t_2)} \sin(\omega t_1) \cos(\omega t_2)\,.
\end{split}
\end{equation}

Expression \eqref{k} implies that the variance $\var S(\omega,\TT)=\overline{S(\omega,\TT)^2}-\overline{S(\omega,\TT)}^2$ of a single-trajectory PSD obeys
\begin{align}
\label{m}
\var S(\omega,T) = 2 \overline{S_\cc(\omega,\TT)}^2 + 2 \overline{S_\ss(\omega,\TT)}^2 + 4 \overline{S_{\cs}(\omega,\TT)}^2\,.
\end{align}
Our first goal is to prove that $\var S(\omega,\TT) \geq \mu^2(\omega,\TT)$,  i.e.,  that $\gamma \geq 1$.
To this end, we notice that
\begin{align}
\label{m1}
\mu(\omega,\TT) = \overline{S_\cc(\omega,\TT)} + \overline{S_\ss(\omega,\TT)} \,,
\end{align}
and rewrite formally eq.~\eqref{m} as
\begin{align}
\label{var}
\var S(\omega,\TT) = \mu^2(\omega,\TT) + \Delta(\omega,\TT) \,,
\end{align}
where
\begin{align}
\label{del}
\Delta(\omega,\TT) = \left(\overline{S_\cc(\omega,\TT)} - \overline{S_\ss(\omega,\TT)}\right)^2  + 4 \overline{S_{\cs}(\omega,\TT)}^2 \,.
\end{align}
The lower bound follows by merely noticing that $\Delta(\omega,\TT) > 0 $ for any $\omega$ and $\TT$. It may be instructive to note that despite the simplicity of its derivation, 
this bound has strong implications: namely, it
shows that fluctuations of $S(\omega,\TT)$ for arbitrary centered Gaussian processes 
exceed generically the mean value $\mu(\omega,\TT)$. This implies  that the knowledge of $\mu(\omega,\TT)$ alone may not be sufficient to fully characterise the behaviour of this random variable. 
It also implies that estimating $\mu(\omega,\TT)$ from numerical or experimental data 
reliably well may require quite large statistical samples.  

In order to prove the upper bound $\gamma \leq \sqrt{2}$, we have to show that $\Delta(\omega,\TT) \leq \mu^2(\omega,\TT)$ or, equivalently, that
\begin{align}
\overline{S_{\sc}(\omega,\TT)}^2 \leq \overline{S_{\cc}(\omega,\TT)} \, \overline{S_{\ss}(\omega,\TT)} \,.
\end{align}
We rewrite next the latter inequality in the explicit form
\begin{equation}
\label{ineq1}
\begin{split}
&\left(\frac{1}{\TT} \int^{\TT}_0 \D t_1 \int^{\TT}_0 \D t_2 \; \overline{X(t_1) X(t_2)} \, \cos(\omega t_1) \sin(\omega t_2)\right)^2 \\
&\qquad \leq  \left(\frac{1}{\TT}  \int^{\TT}_0 \D t_1 \int^{\TT}_0 \D t_2 \; \overline{X(t_1) X(t_2)} \, \cos(\omega t_1) \cos(\omega t_2)\right) \\
&\qquad \qquad {}\times \left(\frac{1}{\TT} \int^{\TT}_0 \D t_1 \int^{\TT}_0 \D t_2 \; \overline{X(t_1) X(t_2)} \, \sin(\omega t_1) \sin(\omega t_2)\right) \,,
\end{split}
\end{equation}
and take advantage of Mercer's theorem \cite{mercer09,mercer}. This theorem asserts that for an arbitrary symmetric, continuous non-negative kernel function $\overline{X(t_1) X(t_2)}$ there exists an orthonormal set of eigenfunctions  $e_k(t)$ defined in $[0,\TT]$ with positive eigenvalues $\lambda_k$, such that the covariance function $\overline{X(t_1) X(t_2)}$ can be expressed by 
\begin{align}
\label{exp}
\overline{X(t_1) X(t_2)} = \sum_{k=1}^{\infty} \lambda_k \, e_k(t_1) \, e_k(t_2)  \,,
 \end{align}
and that this series converges absolutely and uniformly.
The eigenvalues $\lambda_k$ and eigenfunctions $e_k(t)$ are generally 
found by solving the homogeneous Fredholm integral equation of the second kind:\footnote{In particular, when $X(t)$ is the standard Brownian motion, $e_k(t) \propto \sin((k-1/2) \pi t/\TT)$ and $\lambda_k = 1/(\pi^2 (k-1/2)^2)$. This corresponds to the celebrated Wiener representation of the BM. For the FBM the eigenfunctions $e_k(t)$ are combinations of the sine and cosine functions as above, but the eigenvalues $\lambda_k$ are not simply multiples of $\pi^2$ and are expressed through the zeros of the Bessel functions $J_H(x)$ and $J_{1-H}(x)$ (see~\cite{kacha}).}
\begin{align}
\label{exp3}
\int^\TT_0 \D t_1\;\overline{X(t_1) X(t_2)} \, e_k(t_1) = \lambda_k \, e_k(t_2) \,.
\end{align}
We substitute the expansion \eqref{exp} in the inequality \eqref{ineq} and introduce the notation
\begin{equation}
\begin{split}
\label{cs}
c_k &= c_k(\omega,\TT) = \sqrt{\frac{\lambda_k}{\TT}} \int^\TT_0 \D t \; e_k(t) \, \cos(\omega t)  \, , \\
 s_k &= s_k(\omega,\TT) = \sqrt{\frac{\lambda_k}{\TT}} \int^{\TT}_0 \D t \; e_k(t) \, \sin(\omega t) \,.
\end{split}
\end{equation}
We can then cast the upper bound in inequality \eqref{ineq} into the form
\begin{align}
\label{cc}
\left(\sum_{k=1}^{\infty} c_k s_k \right)^2 \leq \left(\sum_{k=1}^{\infty} c^2_k\right) \left(\sum_{k=1}^{\infty} s^2_k\right) \,,
\end{align}
where the existence of the limits on the right- and left-hand-sides is ensured by the absolute convergence of the expansion \eqref{exp}. Equation \eqref{cc} is the standard Cauchy-Schwarz inequality, which thus proves the upper bound on the coefficient of variation $\gamma$. Note that the inequality 
becomes an  identity for $\omega=0$ when both sides vanish.

\section{The moment-generating function}
\label{2}

In order to evaluate the probability distribution function of $S(\omega,\TT)$, we consider the moment-generating function $\Phi_{\kappa}$ of the single-trajectory PSD of an arbitrary real-valued, centered Gaussian process $X(t)$. This function is defined by
\begin{align}
\Phi_{\kappa} = \overline{\exp\left(- \kappa \, S(\omega,\TT)\right)} \,, \qquad \kappa \geq 0 \,.
\end{align}
We rewrite eq.~\eqref{def1} in the form
\begin{equation}
\label{S}
\begin{split}
S(\omega,\TT) &=  
 \frac{1}{\TT}\int^\TT_0 \int^\TT_0 \D t_1\, \D t_2 \; X(t_1) \, X(t_2)\, \cos\left(\omega (t_1 - t_2)\right) \\
 &= \frac{1}{\TT} \int^\TT_0 \int^\TT_0 \D t_1 \D t_2 \; X(t_1)\, X(t_2) \Big(\cos(\omega t_1) \cos(\omega t_2) + \sin(\omega t_1) \sin(\omega t_2)\Big) \\ 
 &= 
\frac{1}{\TT} \left(\int^\TT_0 \D t \; X(t) \cos(\omega t)\right)^2 + \frac{1}{\TT} \left(\int^\TT_0 \D t \; X(t) \sin(\omega t)\right)^2 \,.
\end{split}
\end{equation}
The expression in the last line ensures that $S(\omega,\TT) \geq 0$ for any $\omega$ and $\TT$. 

We take advantage of the Karhunen-Loeve decomposition (see, e.g., \cite{KL}),  according to which any zero-mean square-integrable Gaussian stochastic process $X(t)$, defined on the interval $[0,\TT]$, admits  the following representation:
\begin{align}
\label{kl}
X(t) = \sum_{k=1}^{\infty} Z_k e_k(t) \,.
\end{align}
Here $e_k(t)$ are the above defined orthonormal eigenfunctions and $Z_k$ are independent, normally distributed random variables,  
with zero mean and variance $\lambda_k$. Substituting \eqref{kl} in  eq.~\eqref{S}, we obtain
\begin{align}
S(\omega,\TT) = \left(\sum_{k=1}^{\infty} \frac{Z_k c_k}{\sqrt{\lambda_k}}\right)^2 + \left(\sum_{k=1}^{\infty} \frac{Z_k s_k}{\sqrt{\lambda_k}}\right)^2,
\end{align}
where the $c_k$ and $s_k$ are defined in eqs.~\eqref{cs}.

Further on, we use the identity
\begin{align}
\E^{-\kappa Y^2} = \frac{1}{2 \sqrt{\pi \kappa}} \int^{\infty}_{-\infty} \D z \; \E^{-z^2/(4 \kappa) + \I  z Y} \, ,
\end{align} 
which permits us to write down $\Phi_{\kappa}$ as the following two-fold integral:
\begin{align}
\label{mm}
\Phi_{\kappa} &= \frac{1}{4 \pi \kappa} \int^{\infty}_{-\infty} \int^{\infty}_{-\infty} \D z_1\, \D z_2\; \exp\left(- \frac{z_1^2 + z_2^2}{4 \kappa}\right) \overline{\exp\left(\I \sum_{k=1}^\infty \frac{Z_k}{\sqrt{\lambda_k}} \Big(z_1 c_k + z_2 s_k\Big)\right)}  \,.
\end{align}
The averaging can now be straightforwardly performed to give
\begin{equation}
\label{m3}
\begin{split}
\Phi_{\kappa} &= \frac{1}{4 \pi \kappa} \int^{\infty}_{-\infty} \int^{\infty}_{-\infty} \D z_1\, \D z_2\; \exp\left(- \frac{z_1^2 + z_2^2}{4 \kappa} - \frac{1}{2} \sum_{k=1}^{\infty} \left(z_1 c_k + z_2 s_k\right)^2\right) \\
&= \left[1 + 2 \sum_{k=1}^{\infty} \left(c_k^2 + s_k^2\right) \, \kappa  + 4  \left(\sum_{k=1}^{\infty} \sum_{p=1}^{\infty} c_k^2 s_p^2 - \left(\sum_{k=1}^{\infty} c_k s_k\right)^2 \right)  \kappa^2
\right]^{-1/2} \,,
\end{split}
\end{equation}
where the coefficient in front of $\kappa^2$ is evidently positive, by virtue of eq.~\eqref{cc}.

The last step consists in identifying the coefficients in front of $\kappa$ and $\kappa^2$ in the last line in eq.~\eqref{m3}. Using eqs.~\eqref{m} and \eqref{m1}, we readily obtain that
\begin{equation}
\begin{split}
\sum_{k=1}^{\infty} \left(c_k^2 + s_k^2\right)  &= \mu(\omega,\TT) \,, \\
 4  \left(\sum_{k=1}^{\infty} \sum_{p=1}^{\infty} c_k^2 s_p^2 - \left(\sum_{k=1}^{\infty} c_k s_k\right)^2 \right) &= 2 \mu^2(\omega,\TT) - \var S(\omega,\TT)\\
 & = \left(2-\gamma^2(\omega,\TT)\right)\mu^2(\omega,\TT) \,.
\end{split}
\end{equation}
By inverting the Laplace transform in eq.~\eqref{m3} we obtain eq.~\eqref{prob}. We note moreover that,
similarly to the parental Gaussian process $X(t)$, the probability density function in eq.~\eqref{prob} is entirely defined by the first two moments of $S(\omega,\TT)$.

\section{Limiting cases and typical behaviour of the PSD}
\label{4}

We consider here two limiting situations in which the functional form of the probability density function in eq. \eqref{prob} simplifies; namely, when $\gamma \to 1$ (i.e., $\var S(\omega,\TT) \to \mu^2(\omega,\TT)$) or when $\gamma \to \sqrt{2}$ (i.e., $\var S(\omega,\TT) \to 2 \mu^2(\omega,\TT)$).  Note that, in general, the coefficient of variation is an oscillating function of the frequency  for fixed $\TT$,  but attains a constant value in the $\TT \to \infty$ limit. In particular, we have  $\gamma \to 1$  in the limit $\TT \to \infty$ for, e.g.,  the sub-diffusive fractional Brownian motion \cite{krapf2} or for the Brownian gyrator model \cite{sara}. In turn, $\gamma \to \sqrt{2}$ holds in the same limit for the super-diffusive fractional Brownian motion \cite{krapf2}.  
Moreover, $\gamma = \sqrt{2}$ holds as an identity for any $\TT$ in any centered Gaussian process when $\omega = 0$, a relation that follows immediately from eqs.~\eqref{var} and~\eqref{del}. 
Indeed, $S(\omega=0,\TT)$ is equal to the squared area under the random Gaussian curve $X(t)$, divided by the observation time (see eq.~\eqref{def1}).  The limiting forms of the probability density function can be conveniently studied using the expression~\eqref{m3}.

When $\gamma = 1$, the expression in the last line in eq.~\eqref{m3} becomes a full square, and thus one has
\begin{align}
\Phi_{\kappa} = \frac{1}{1 +\kappa\, \mu(\omega,\TT)} \,,
\end{align}
implying that the probability density function is a simple exponential. This form has been experimentally  verified for the sub-diffusive fractional Brownian motion \cite{krapf2} and for the Brownian gyrator model \cite{sara}.

When $\gamma = \sqrt{2}$, the coefficient in front of $\kappa^2$ in eq.~\eqref{m3} vanishes and the moment-generating function becomes
\begin{align}
\Phi_{\kappa} = \frac{1}{\sqrt{1 + 2 \mu(\omega,\TT) \kappa}} \,,
\end{align}
which signifies that the probability density function converges to the $\chi^2$-distribution with one degree of freedom, i.e.,
\begin{align}
\label{z1}
P(s) = \frac{1}{\sqrt{2 \pi \mu(\omega,\TT) s}} \exp\left(- \frac{s}{2 \mu(\omega,\TT)}\right) \,,\qquad s>0.
\end{align}
Note that for $\omega =0$, when the single-trajectory PSD defines the \textit{squared} area under random curve $X(t)$, this result simply states that the area itself has a Gaussian distribution, which is an \textit{a priori} known result. The form in \eqref{z1} has also been verified in experimental analyses of fractional Brownian motion processes \cite{krapf2} and for the Brownian gyrator model \cite{sara}.

Since $\gamma \geq 1$, the magnitude of fluctuations of $S(\omega,\TT)$ exceeds generically its mean value. In other words, the fluctuations of $S(\omega,\TT)$ over different realizations of $X(t)$ are significant and
$\mu(\omega,\TT)$ is most likely dominated by some \textit{atypical} realisations of the random process $X(t)$.
This means, in turn, that in order to correctly reproduce the analytical predictions from numerics or experiments, the number of realizations of the process has to be large enough in order to ``catch''  such rare trajectories. 
The function in eq.~\eqref{prob} attains its maximal value at $s=0$, so that we can expect that typical trajectories most often yield a value of $S(\omega,\TT)$ smaller than the average. In our case, we estimate the typical value of $S(\omega,\TT)$ as the exponential of the average of the logarithm of $S(\omega,\TT)$:
\begin{equation}
\mu_{\typ}(\omega,\TT)=\exp\left[\overline{\ln S(\omega,\TT)}\right]\,.
\end{equation}  
For the probability density function given in eq.~\eqref{prob} we can perform the corresponding integral exactly, and obtain
 \begin{align}
\overline{\ln\left(\frac{S(\omega,\TT)}{\mu(\omega,\TT)}\right)} = \ln\left(\frac{1+ \sqrt{2 - \gamma^2}}{2}\right) - C \,,
\end{align}   
where $C \approx 0.577$ is the Euler-Mascheroni constant. (Notice that the convergence of the integral is guaranteed by the double-sided inequality~\eqref{ineq}.) Consequently, the typical value of a single-trajectory PSD is given by
\begin{equation}
\label{typ}
\begin{split}
\mu_{\typ}(\omega,\TT) &= \mu(\omega,\TT)\, \E^{-C} \left(\frac{1 + \sqrt{2 - \gamma^2}}{2}\right)  \,.
\end{split} 
\end{equation}
As intuitively expected, $\mu_{\typ}(\omega,\TT)$ is smaller than $\mu(\omega,\TT)$ and the difference between them is more pronounced for $\gamma$ close to $\sqrt{2}$ than for $\gamma$ close to~$1$. Therefore, in order to obtain reliable estimates of $\mu(\omega,\TT)$, statistical samples need to be larger in the former than in the latter case.

\section{Conclusions}
\label{3}

Summarizing, we have analysed here the statistical properties  
of the fluctuations of a single-trajectory power spectral density $S(\omega,\TT)$ of centered Gaussian processes $X(t)$, going beyond its first moment, which is the main focus of most analyses. We have presented a formal proof of the statement that for an arbitrary Gaussian process the noise-to-signal ratio $\gamma$, defined as the ratio of the standard deviation and the mean value of $S(\omega,\TT)$, obeys a double-sided inequality $1 \leq \gamma \leq \sqrt{2}$ for any $\omega$ and any $\TT$. The bound $\gamma>1$ implies that the magnitude of fluctuations generically exceeds the mean value, meaning that the realisation-to-realisation fluctuations are very significant. 
Using the Karhunen-Loeve decomposition of $X(t)$, we evaluated the full probability density function of $S(\omega,\TT)$, which holds for arbitrary centered Gaussian processes and for any value of the frequency and of the observation time. Finally, we have discussed the typical behaviour of a single-trajectory power spectral density, which is most likely to be observed for small statistical ensembles of trajectories.

\section*{Acknowledgments}
The authors acknowledge helpful discussions with Sergio Ciliberto. LP is grateful to the LPTMC, Sorbonne Université, for hospitality during the redaction of this work. AS acknowledges FWF Der Wissenschaftsfonds for funding through the Lise-Meitner Fellowship (Grant No.\ M 3300-N). LR acknowledges the support of Italian National Group of Mathematical Physics (GNFM) of INDAM, and of Ministero dell’Istruzione e dell’Università e della Ricerca (MIUR), Italy, Grant No.\ E11G18000350001 ``Dipartimenti di Eccellenza 2018–2022''.

\end{document}